\documentclass[aps,prl,twocolumn,groupedaddress,superscriptaddress,showpacs]{revtex4}

\usepackage{graphicx}
\usepackage{colortbl}

\makeatletter
\def\btt#1{\texttt{\@backslashchar#1}}%
\DeclareRobustCommand\bblash{\btt{\@backslashchar}}%
\makeatother

\begin{document}


\title{Dynamics of Multiferroic Domain Wall in Spin-Cycloidal Ferroelectric DyMnO$_{3}$} 

\author{F. Kagawa}
\affiliation{Multiferroics Project, ERATO, Japan Science and Technology Agency (JST), c/o Department of Applied Physics, University of Tokyo, Tokyo 113-8656, Japan}

\author{M. Mochizuki}
\affiliation{Multiferroics Project, ERATO, Japan Science and Technology Agency (JST), c/o Department of Applied Physics, University of Tokyo, Tokyo 113-8656, Japan}

\author{Y. Onose}
\affiliation{Multiferroics Project, ERATO, Japan Science and Technology Agency (JST), c/o Department of Applied Physics, University of Tokyo, Tokyo 113-8656, Japan}
\affiliation{Department of Applied Physics, University of Tokyo, Tokyo 113-8656, Japan}

\author{H. Murakawa}
\affiliation{Multiferroics Project, ERATO, Japan Science and Technology Agency (JST), c/o Department of Applied Physics, University of Tokyo, Tokyo 113-8656, Japan}

\author{Y. Kaneko}
\affiliation{Multiferroics Project, ERATO, Japan Science and Technology Agency (JST), c/o Department of Applied Physics, University of Tokyo, Tokyo 113-8656, Japan}

\author{N. Furukawa}
\affiliation{Multiferroics Project, ERATO, Japan Science and Technology Agency (JST), c/o Department of Applied Physics, University of Tokyo, Tokyo 113-8656, Japan}
\affiliation{Department of Physics and Mathmatics, Aoyama Gakuin University, Kanagawa 229-8558, Japan}

\author{Y. Tokura}
\affiliation{Multiferroics Project, ERATO, Japan Science and Technology Agency (JST), c/o Department of Applied Physics, University of Tokyo, Tokyo 113-8656, Japan}
\affiliation{Department of Applied Physics, University of Tokyo, Tokyo 113-8656, Japan}
\affiliation{Cross-Correlated Materials Research Group (CMRG), ASI, RIKEN, Wako 351-0198, Japan}

\date{\today}

\begin{abstract}
	We report the dielectric dispersion of the giant magnetocapacitance (GMC) in multiferroic DyMnO$_{3}$ over a wide frequency range.
	The GMC is found to be attributable not to the softened electromagnon but to the electric-field-driven 
motion of multiferroic domain wall (DW). 
	In contrast to conventional ferroelectric DWs, the present multiferroic DW motion holds extremely high relaxation rate of $\sim$$10^{7}$ s$^{-1}$ even at low temperatures.
	This mobile nature as well as the model simulation suggests that the multiferroic DW is not atomically thin as in ferroelectrics but thick,
reflecting its magnetic origin.

\end{abstract}


\maketitle

	Prospective materials toward electric control of magnetism in a solid are multiferroics, in which electric and magnetic orders coexist \cite{Fiebig2005,Cheong2007}. In conventional ferromagnetic or ferroelectric materials, the motion of ferroic domain walls (DWs) is a key to the functions; it provides huge linear-response (i.e., magnetic or dielectric susceptibility) \cite{Kittel1949,Book1,Nakamura1988} as well as low-field control of the ferroic domain. 
	By analogy, the control of the multiferroic [i.e. concurrently (anti)ferromagnetic and ferroelectric] DWs in multiferroics may provide a new prospective means to attain the electric (magnetic) control of the magnetic (ferroelectric) domain on a macroscopic scale as well as the enhancement of the dynamical magnetoelectric (ME) susceptibility.
	In this Letter, through dielectric measurements over a wide frequency region on the multiferroic perovskite, DyMnO$_{3}$ \cite{GotoPRL}, we identify the microscopic motion of multiferroic DW as a source of large magnetoelectric coupling. 
	The multiferroic DW is found to be dynamically active even at low temperatures, enabling the electric control of magnetic domains via the macroscopic DW movement.
	These characteristics suggest that the present multiferroic DW is thick 
in width, in contrast to the conventional thin ferroelectric DW.
%

\begin{figure}
\includegraphics[width=8.5cm,height=9.6cm]{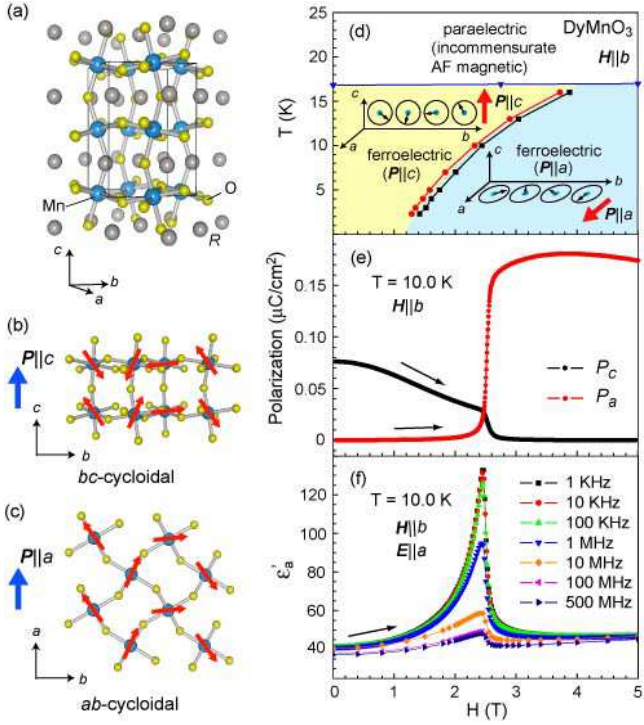}
\caption{\label{Fig1}(Color online) 
	(a) Crystal structure of DyMnO$_{3}$. 
	(b) and (c) Magnetic structures of $bc$-cycloidal spin order (b) and $ab$-cycloidal spin order (c). 
	(d) Magnetic field-temperature phase diagram of DyMnO$_{3}$ for the case of {\boldmath $H$}$\parallel$$b$. 
	The phase boundaries were determined from the dielectric measurements. 
	The closed squares and circles represent the transition points in increasing and decreasing field processes, respectively. 
	(e) Magnetic-field dependence ({\boldmath $H$}$\parallel$$b$) of polarization at 10 K along the $c$ axis ({\boldmath $P$}$\parallel$$c$) and the $a$ axis ({\boldmath $P$}$\parallel$$a$). 
	(f) Magnetic-field dependence ({\boldmath $H$}$\parallel$$b$) of dielectric constant at 10 K along the $a$ axis at various frequencies.}

\end{figure}

	In multiferroic materials, ferroelectric polarization {\boldmath $P$} is often induced by magnetic order 
through the inverse Dzyaloshinskii-Moriya (DM) mechanism \cite{inverseDM}.
	In the perovskite manganites $R$MnO$_{3}$ ($R$ = Tb and Dy) with GdFeO$_{3}$-type distortion [Fig. 1(a)], there is growing evidence that the cycloidal magnetic order induces the ferroelectricity through the inverse DM mechanism \cite{KenzelmannPRL,ArimaPRL}, i.e., {\boldmath $P$} $\propto$ $\sum_{i,j}^{}$ {\boldmath $e$}$_{ij}$ $\times$ ({\boldmath $S$}$_{i}$ $\times$ {\boldmath $S$}$_{j}$), where {\boldmath $S$}$_{i}$ ({\boldmath $S$}$_{j}$) is the electron spin vector at site $i$ ($j$) and {\boldmath $e$}$_{ij}$ is the unit vector connecting the two sites. 
	In TbMnO$_{3}$ or DyMnO$_{3}$, the $bc$-cycloidal spin order is emergent at low temperatures with the propagation vector {\boldmath $q$} along the $b$ axis [{\boldmath $q$}$\parallel$${b}$, see Fig. 1(b)] 
\cite{GotoPRL,KenzelmannPRL,KimuraPRB}; thus {\boldmath $P$} is induced along the $c$ axis. 
	It is also confirmed that a magnetic field {\boldmath $H$}$\parallel$${b}$ induces the polarization flop from {\boldmath $P$}$\parallel$${c}$ to {\boldmath $P$}$\parallel$${a}$ with the direction of {\boldmath $q$} ($\parallel$${b}$) unchanged \cite{KimuraPRB,KimuraNature,AliouanePRB,StrempferPRB}, as exemplified in Figs. 1(d) and 1(e) for DyMnO$_{3}$. This behavior is attributable to the flop of the spin cycloid plane from $bc$ to $ab$ [Fig. 1(c)]. 
	The salient ME feature characteristic of DyMnO$_{3}$, which is investigated here, is the giant magnetocapacitance (GMC) emerging concomitantly with the magnetic-field-induced $P$ flop \cite{GotoPRL}: as shown in Fig. 1(f), the dielectric constant ({\boldmath $E$}$\parallel$${a}$) shows a large enhancement, from  $\sim$40 to $\sim$130 (in the case of 10 K), in the course of the magnetic-field-induced transition. This phenomenon may give an important clue to the general strategy to realize the colossal ME effect.

	We have investigated the magnetocapacitance with varying frequencies of electric field ({\boldmath $E$}$\parallel$${a}$) from 1 KHz to 500 MHz using two equipments: a LCR meter (Agilent E4980A) for 1 KHz-2 MHz and an impedance analyzer (Agilent E4991A) for 1 MHz-500 MHz.
	Figure 1(f) displays the magnetic-field dependence of the real part of dielectric constant, $\epsilon$$_{a}$$\rq$ ({\boldmath $E$}$\parallel$$a$). 
	Note that the GMC emerges below 100 KHz but is considerably suppressed above 10 MHz. 
	This behavior clearly demonstrates that the GMC has prominent dielectric dispersion around 1 MHz at 10 K. 
	More insight into the origin of GMC is obtained from the spectral analysis. We measured the spectra of real and imaginary parts of dielectric constant, $\epsilon$$_{a}$$\rq$ and $\epsilon$$_{a}$$\rq$$\rq$ ({\boldmath $E$}$\parallel$$a$), at various magnetic fields and temperatures. As a typical example, the results at 10 K are shown in Figs. 2(a) and 2(b). 
	These spectra ensure that the GMC is a phenomenon emergent only below 10$^{5}$-10$^{6}$ Hz. 
	A new aspect revealed here is that the spectral shape of GMC is not the resonance type but the relaxation type, indicating that the GMC does not arise from bosonic excitations, such as soft-mode phonons \cite{BarkerPR} and electrically active magnons (electromagnons) \cite{PimenovNatPhys,KatsuraPRL_electromagnon}. 
	This finding is clearly incompatible with the prevailing argument that the softening of electromagnon results in the GMC at low frequencies \cite{KatsuraPRL_electromagnon,StrempferPRB}. 
	The origin of dielectric relaxation can be deduced by considering several features: (i) the magnetic-field dependence of the magnitude of relaxation mode, $\Delta$$\epsilon$$_{a}$$\rq$, (ii) the temperature dependence of the relaxation rate at the flop transition, and (iii) the value of the relaxation rate itself. 
	Here we evaluated $\Delta$$\epsilon$$_{a}$$\rq$($H, T$) by $\epsilon$$_{a}$$\rq$($H$, $T$, 1KHz) - $\epsilon$$_{a}$$\rq$($H$, $T$, 500MHz) and defined the relaxation rate as 1/$\tau$($H, T$) = 2$\pi$$f_{\rm peak}$($H, T$), 
where $f_{\rm peak}$ is the frequency of $\epsilon$$_{a}$$\rq$$\rq$ peak. 
	Below we refer to (i)-(iii) and conclude that the GMC is provided by the motion of multiferroic DWs between the {\boldmath $P$}$\parallel$${\pm a}$ ($bc$-cycloidal) and the {\boldmath $P$}$\parallel$${\pm c}$ ($ab$-cycloidal) domains. Hereafter, the relevant multiferroic DWs are abbreviated as DW$_{\pm a/\pm c}$ [see Fig. 3(a)].

\begin{figure}
\includegraphics[width=8.5cm,height=7.7cm]{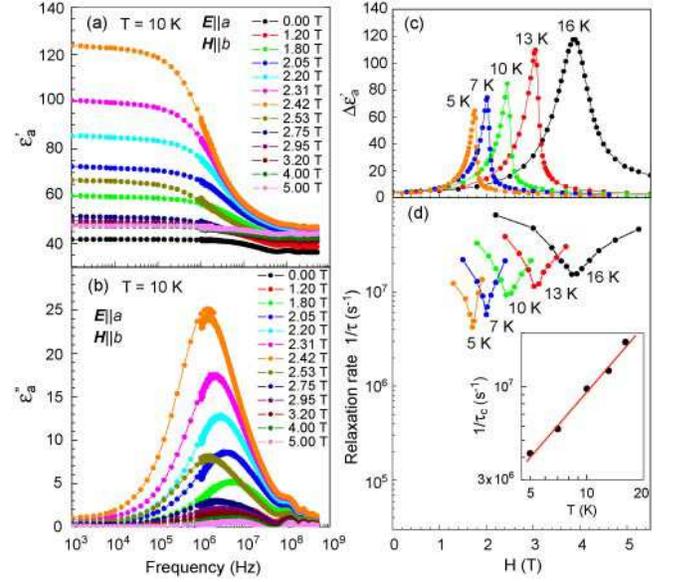}
\caption{\label{Fig2}(Color online) 
	(a) and (b) Spectra of dielectric constant at 10 K under various magnetic fields: real part (a) and imaginary part (b). 
	(c) and (d) Magnetic-field dependence of the magnitude of the relaxation mode (c) and the relaxation rate (d) at given temperatures. 
	Inset in (d): Temperature dependence of the relaxation rate at the flop transition, 1/$\tau$$_{\rm c}$ (double logarithmic plot).
	The solid line is a guide to the eye.}

\end{figure}

	The magnetic-field dependence of $\Delta$$\epsilon$$_{a}$$\rq$ and 1/$\tau$ at various temperatures are shown in Figs. 2(c) and 2(d), respectively. Note that $\Delta$$\epsilon$$_{a}$$\rq$ shows a sharp peak even at 5 K. 
	This tendency is not expected for the case that the GMC were due to the polarization fluctuations; at low temperatures the polarization flop transition is of the first order and thus the enhancement of fluctuations is unlikely. 
	In Fig. 2(d), we focus on the local minimum value of 1/$\tau$($H$) at a given temperature, say 1/$\tau$$_{\rm c}$ 
(i.e., 1/$\tau$ at the polarization flop transition).
	Note that although the 1/$\tau$ relevant to polarization fluctuations should become larger toward low temperatures in general \cite{OzakiJPSJ,Book2}, the currently observed 1/$\tau$$_{\rm c}$ rather diminishes. 
	Therefore, in terms of (i) and (ii), it is unlikely that the observed relaxation comes from the polarization fluctuations. 
	As the origin of such a relaxation characteristic, the microscopic displacement of DW under oscillating $E$ is the most plausible. 
%

\begin{figure}
\includegraphics[width=8.5cm,height=10.4cm]{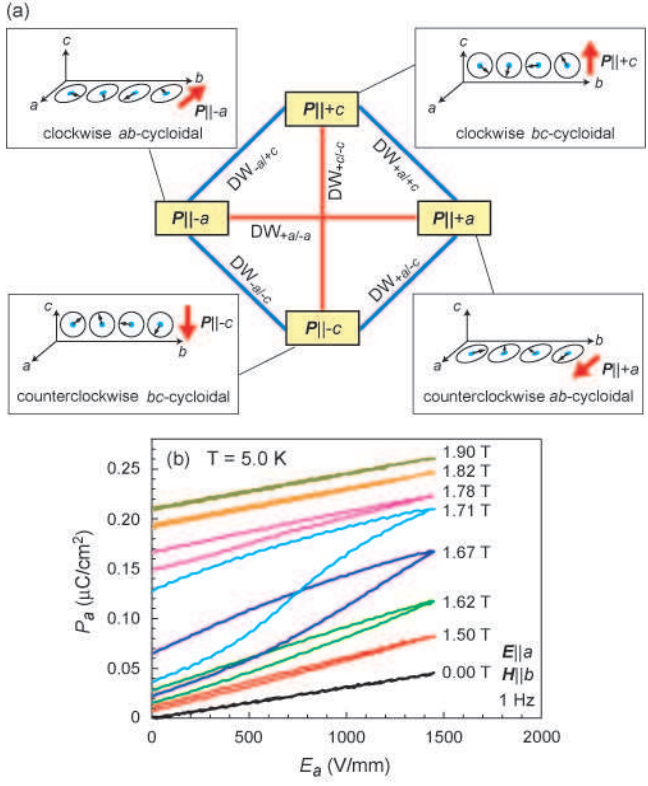}
\caption{\label{Fig3} (Color online) 
	(a) Various multiferroic DWs conceivable in DyMnO$_{3}$. 
	The red lines represent the DW through which both the spin helicity ({\boldmath $S$}$_{i}$ $\times$ {\boldmath $S$}$_{j}$) and the polarization ($P$) are reversed, while the blue lines represent the DW through which they rotate by 90 degrees. 
	(b) Initial $P$-$E$ curves ({\boldmath $E$}$\parallel$$a$) under various magnetic fields at 5 K,
demonstrating electric-field-induced movement of multiferroic DW between the {\boldmath $P$}$\parallel$$c$ ($bc$-cycloidal) and the {\boldmath $P$}$\parallel$$a$ ($ab$-cycloidal) domains.}

\end{figure}

	Among the various possible multiferroic DWs in DyMnO$_{3}$ [Fig. 3(a)], the relevant ones that may give rise to large $\epsilon$$_{a}$$\rq$ response are the motion of DW$_{+a/-a}$ and/or DW$_{\pm a/\pm c}$.
	In the {\boldmath $P$}$\parallel$$a$ phase, we can identify the DW$_{+a/-a}$ contribution to $\epsilon$$_{a}$$\rq$ from the difference between $\epsilon$$_{a}$$\rq$ in
{\boldmath $P$}$\parallel$$\pm a$ multi-domain state and $\epsilon$$_{a}$$\rq$ in {\boldmath $P$}$\parallel$$+a$ single-domain state.
	We found that although the DW$_{+a/-a}$ contribution is quite small ($\Delta$$\epsilon$$_{a}$$\rq$ $<$ 7), the DW$_{+a/-a}$ motion also shows the relaxation-type spectra and its 1/$\tau$ 
is comparable to 1/$\tau$$_{c}$ of GMC (not shown). 
	Therefore, from the quantitative point of view, the GMC emerging only near the flop transition is 
also attributable to one of the DW motions, namely the DW$_{\pm a/\pm c}$ motion \cite{PolingEffect}.
	Remarkably, this DW$_{\pm a/\pm c}$ motion holds a high relaxation rate of $\sim$ 10$^{7}$ s$^{-1}$ even at low temperatures, in contrast to the conventional ferroelectric DW motion, which usually freezes at low temperatures \cite{HuangPRB}. 
	In fact, the temperature dependence of 1/$\tau$$_{\rm c}$ obeys a power-law rather than the Arrhenius law [the inset of Fig. 2(d)], resulting in the gradual slowing down of 1/$\tau$$_{\rm c}$ toward low temperatures and thus in the high relaxation rate even at 5 K.
%

\begin{figure}
\includegraphics[width=8.5cm,height=5.0cm]{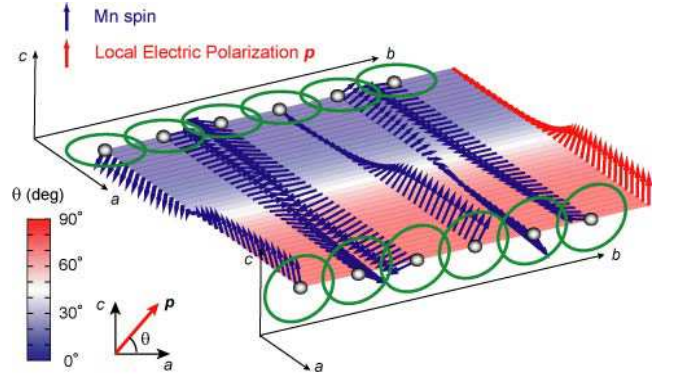}
\caption{\label{Fig4}(Color online) 
	Calculated DW structure between the {\boldmath $P$}$\parallel$$+c$ ($bc$-cycloidal) and {\boldmath $P$}$\parallel$$+a$ ($ab$-cycloidal) domains for 36 $\times$ 6 Mn sites. 
	Blue and red arrows represent the Mn spins and the local electric polarizations, respectively. 
	The color gradation represents the angle of local electric polarization relative to the $a$ axis. The angle becomes 45 degrees along the DW center, which runs parallel to the $b$ axis.}

\end{figure}

	Since the DW$_{\pm a/\pm c}$ motion is still unfrozen even at the lowest temperature, 5 K, the macroscopic movement of depinned DW$_{\pm a/\pm c}$ is expected under a strong electric field. 
	In conventional ferroelectrics, if one attempted to induce the macroscopic movement of DW at 5 K, 
an impractically strong electric field or long waiting time would be required because the DW motion via nucleation on the DW is hard to occur at such low temperatures \cite{MillerPR1960,ShinNature}. 
	In DyMnO$_{3}$, however, the macroscopic DW$_{\pm a/\pm c}$ movement does occur even at 5 K;
figure 3(b) displays the "initial" $P$-$E$ curve ({\boldmath $E$}$\parallel$$a$) under various magnetic fields. 
	Note that the hysteresis behavior is observed only near the flop transition field ($\sim$ 1.71 T at 5 K). 
	This hysteresis indicates that the initial and final states are different. 
	Since we took the experimental procedure to suppress the mixture of {\boldmath $P$}$\parallel$-$a$ domain in the {\boldmath $P$}$\parallel$$a$ phase \cite{experiment}, this irreversible increase in $P_{a}$ can be regarded as the consequence of the macroscopic movement of DW$_{+a/+c}$ and/or DW$_{+a/-c}$. 
	Therefore, the irreversible $P$-$E$ curve suggests that the magnetic cycloidal plane was changed from $bc$ to $ab$ by an electric stimulation through the DW movement. 

	The fact that the DW$_{\pm a/\pm c}$ motion does not freeze even at 5 K implies that the DW$_{\pm a/\pm c}$ motion 
does not occur via the nucleation process on the DW \cite{MillerPR1960, ShinNature}. 
	Moreover, it is known that there is a close relationship between the DW mobility and the DW thickness
\cite{ChoudhuryJAP}.
	In this context, the DW$_{\pm a/\pm c}$ structure is of particular interest and importance.
	Here we postulate that the DW structure is determined by minimizing 
mainly the magnetic energy cost, because the ferroelectricity of magnetic origin in DyMnO$_{3}$ is quite weak in magnitude
compared with conventional ferroelectrics.
	To calculate the internal structure of DW$_{+a/+c}$, we employed the two-dimensional classical $J_{1}$-$J_{2}$ model with ferromagnetic $J_{1}$ (= 0.8 meV) on the tetragonal $x$ and $y$ bonds and antiferromagnetic $J_{2}$ (= 0.96 meV \cite{GontcharJMMM}) on the diagonal bonds along the $b$ axis: these values reproduce a spiral spin order with {\boldmath $q$} = 0.36 ($\parallel$$b$) \cite{GotoPRL}.	
	To mimic the experimental situation where the $ab$- and $bc$-plane spin cycloids are degenerate at the phase boundary, we added the single-ion anisotropy term, $D\sum_{i}^{}$($S_{ai}^{2}+S_{ci}^{2}$) with $D$ = 0.20 meV,
and the orthorhombic anisotropy term, $\beta\sum_{i}^{}$$S_{ai}^{2}S_{ci}^{2}$ with $\beta$ = 0.005 meV \cite{MatsumotoJPSJ}: the former excludes the $ac$-cycloidal order, while the latter generates an energy barrier between $ab$- and $bc$-cycloidal states.
	By minimizing the energy, we obtained a stable structure of DW$_{+a/+c}$.
	Then local electric polarization {\boldmath $p$} was calculated on the basis of the inverse DM mechanism. 
	As shown in Fig. 4, the calculated DW$_{+a/+c}$ structure is thick ($\sim$ 20 unit cells), reflecting the Heisenberg nature of the constituent spins. 
	Within the DW$_{+a/+c}$, the cycloid plane continuously rotates from $ab$ to $bc$ and thus
the local polarization continuously rotates from {\boldmath $p$}$\parallel$$a$ to {\boldmath $p$}$\parallel$$c$.
	The latter feature is reminiscent of the ferromagnetic N\'eel wall in the Heisenberg spin systems.

	The gradual rotation of polarization in the thick multiferroic DW is a remarkable feature, in contrast to the situation in the conventional ferroelectric DWs, which is Ising-like and atomically thin 
\cite{MillerPR1960,ShinNature,PadillaPRB}.
	Note that Rochelle salt, which show the weak polarization comparable to that of DyMnO$_{3}$ \cite{Rochelle}, is expected to exhibit the DW of a few unit cell thickness \cite{RochelleDW,Catalan}. 
	Therefore the thick multiferroic DW is attributable not merely to its small magnitude of polarization but to 
the ferroelectricity of magnetic origin.
	The motion of such Heisenberg-like thick DWs is generally via magnon excitations with a small gap; thus, the magnon excitations and the 1/$\tau$ of thick-DW motion are expected to diminish gradually toward low temperatures because of the small gap. 
	In DyMnO$_{3}$, the gradual slowing down of 1/$\tau$$_{\rm c}$ obeying a power law [the inset of Fig. 2(d)] indicates that a gap, if any, is comparable with or even smaller than the present temperature range, 5-16 K: the gap may originate from the electromagnon-excitation gap at the $P$ flop transition.
	The possible small gap also implies thick DWs. 
	Therefore the present experimental and numerical results are consistent, suggesting that the multiferroic DW$_{\pm a/\pm c}$ is the Heisenberg-like thick DW rather than the Ising-like thin DW.
	This may explain why the present multiferroic DW is well mobile at low temperatures even though it works also as the ferroelectric DW. 
	Finally, on the basis of the scaling behavior found by Catalan \textit{et al}. \cite{Catalan}, 
such thick DW implies large-size multiferroic domains as obserbed in the different kind of 
multiferroics BiFeO$_{3}$ \cite{BiFeO3}.

	In summary, we have investigated the dielectric dispersion of the GMC in DyMnO$_{3}$ over a wide frequency range and found that GMC is attributable not to the softened electromagnon but to the local motion of the multiferroic DW between $bc$-plane spin cycloid ({\boldmath $P$}$\parallel$$c$) and $ab$-plane spin cycloid ({\boldmath $P$}$\parallel$$a$) domains. The relaxation rate of this DW motion holds high ($\sim$10$^{7}$ s$^{-1}$) even at 5 K. 
	These results suggest that the multiferroic DW emerging at the polarization flop transition is the Heisenberg-like 
thick DW in contrast to the Ising-like thin DW in conventional ferroelectrics.
	This difference may explain the mobile multiferroic DW even at low temperatures.

	The authors thank N. Kida, N. Nagaosa, S. Miyahara, and S.
Onoda for fruitful discussion.
	This work was in part supported by Grant-In-Aid for Science Research from the MEXT (Nos. 20046004, 20340086, and 19684011), Japan.


\begin{references}
\bibitem{PolingEffect} Around the first-order flop transition, the phase separation of {\boldmath $P$}$\parallel$$a$ 
and {\boldmath $P$}$\parallel$$c$ phases inevitably occurs; the {\boldmath $P$}$\parallel$$+a$ 
poling under 6.5 T did not get rid of GMC.
\bibitem{experiment} We confirmed that once the {\boldmath $P$}$\parallel$$-a$ domain is suppressed with the field-cooling procedure in the {\boldmath $P$}$\parallel$$a$ phase, {\boldmath $P$}$\parallel$$-a$ domain remains suppressed even if the crystal undergoes the polarization flop transitions to the {\boldmath $P$}$\parallel$$c$ phase and then to the {\boldmath $P$}$\parallel$$a$  phase with sweeping magnetic-field. 
Prior to each $P$-$E$ curve measurement, 
we first prepeared the {\boldmath $P$}$\parallel$$+a$ single-domain state and next the magnetic field was decreased down to 0 T. 
Then the field was increased to the target field and we recorded the initial $P$-$E$ curve to detect the irreversible process of the DW movement.
\end{references}
\end{document}